\documentclass[%
reprint,
 aps,
]{revtex4-1}
\usepackage[fleqn]{amsmath}
\usepackage{graphicx}
\usepackage{dcolumn}
\usepackage{bm}
\usepackage{lmodern}

\usepackage[charter]{mathdesign}
\usepackage{caption}
\usepackage[font=small]{subcaption}
\usepackage{float}

\begin{document}

\preprint{KKZ4}

\title{ON THE NATURE OF PERIODICALLY PULSATING RADIATION SOURCES}

\author{A.V.~Kirichok}
\email{sandyrcs@gmail.com}
\author{V.M.~Kuklin}
\affiliation{V.N.~Karazin Kharkiv National University, Institute for High Technologies\\4 Svobody Sq., Kharkiv 61022, Ukraine}

\author{A.G.~Zagorodny}
\affiliation{Bogolyubov Institute for Theoretical Physics\\14-b, Metrolohichna str., Kiev, 03680, Ukraine}

\date{\today}%

\begin{abstract}
A change in the character of maser generation in a two-level system is found when the initial population inversion exceeds some threshold value proportional to the square root of the total number of atoms. Above this threshold, the number of photons begins to grow exponentially with time and the pulse with short leading edge and broadened trailing edge is generated. In this work, we attempt to explain the nature of this threshold. Coherent pulse duration estimated by its half-width increases significantly with increasing inversion, if all other parameters are fixed and the absorption is neglected. The inclusion of the energy loss of photons leads to the fact that the duration of coherent pulse is almost constant with increasing inversion, at least well away from the threshold. If there is exist a recovery mechanism for the population inversion, the pulsating mode of stimulated emission generation becomes possible. The integral radiation intensity at this may be increased several times. This approach can be used for analysis of the cosmic radiation that might help explain a great variety of pulsating radiation sources in space.
 

\end{abstract}

\keywords{new threshold of stimulated emission, pulse of coherent emission, repetition pulse train, pulsating cosmic radiation}
\maketitle


\section{\label{sec:level1}Introduction}

 Description of physical phenomena based on the systems of partial differential equations, derived from the observations and experimental facts, often conceals from an investigator some essential features, especially in those cases, when the researchers do not expect to find anomalies and qualitative changes in the dynamics of systems in given range of variables and parameters. Namely such a case of unusual behavior of a two-level quantum system was found in attempting to separate a stimulated component from the total radiation flow.

 In the beginning of the past century, A. Einstein has proposed the model of two-level system, which has demonstrated the possibility of generation of both spontaneous and induced (stimulated) emission when the initial population inversion is sufficiently large \cite{einstein1916quantentheorie}. Usually, the term spontaneous emission denotes the emission of oscillator (or other emitter) which not forced by external field of the same frequency. As for other influences on the characteristics of the spontaneous emission, there is nothing to say definitely. Although the dynamics of spontaneous processes usually shows a steady recurrence and invariance, there is evident \cite{goy1983observation} that the characteristics of the spontaneous processes can vary with change of environment. By induced or simulated emission is usually meant the emission produced because of an external field action on the emitting source at the radiation frequency. 

 There were difficulties in the quantum description with interpretation of the stimulated emission as coherent, where in contrast to the classical case it was impossible to say anything about the phases of the fields emitted by individual atoms and molecules. However, C. Townes believed \cite{townes1965production} that ``the energy delivered by the molecular systems has the same field distribution and frequency as the stimulating radiation and hence a constant (possibly zero) phase difference''. 

 If we assume, relying upon the results of the studies of fluctuation correlations in the laser radiation \cite{brown1957interferometry, brown1958interferometry}, that a stimulated emission has a high proportion of the coherent component, one can find a threshold of coherent radiation at a certain critical value of population inversion \cite{kuklin2012realization}. The specific feature of this threshold is that it follows from the condition that the initial value of the population inversion is equal to the square root of the total number of states. On the other hand, the change in the nature of the process near the threshold is evident, even without making any other assumptions. Above this threshold, the number of photons begins to grow exponentially with time. Herewith, below the threshold there no exponential growth.

 It is known that at low levels of spontaneous component and far above the maser generation threshold the number of photons growths exponentially and the radiation is largely a coherent \cite{birnbaum1965optical, blotmbergen1965nonlinear}. The meaningful indicator of the collective character of stimulated emission is the so-called photon degeneracy, which is defined as the average photon number contained in a single mode of optical field (see, for example \cite{he1999physics}). For the incoherent light, this parameter does not exceed unity, but for even the simplest He-Ne maser it reaches the value of $10^{12} $ as was shown in the early works (see \cite{birnbaum1965optical}).  

 It is of interest to go further and analyze the consequences of consideration of the spontaneous emission as a random process (at least, in a homogeneous medium) and induced process as a coherent process. It is clear that the separation of total radiation into two category:  the stimulated -- coherent and spontaneous -- random or incoherent will be idealized simplification. However, such separation may explain, at least qualitatively, the nature of the radiation emitted by two-level quantum system near to exposed threshold. 

 Another indirect proof of the existence of such a threshold is the following observation. The intensity of the spontaneous emission, which is non-synchronized (randomly distributed) over oscillators phases is known to be proportional to their number. The intensity of the coherent stimulated emission is proportional in turn to the square of the number of oscillators. It is easy to see that the exposed threshold corresponds to the case when the intensity of spontaneous and stimulated coherent radiation become equal. 

 In papers \cite{kuklin2012realization, kirichok2013formation}, we have shown that under these conditions the pulse of stimulated radiation with a characteristic profile is formed when the initial population inversion slightly exceeds the threshold. The leading edge of the pulse due to the exponential growth of the field is very sharp due to the exponential growth of the field, and the trailing edge is rather broadened. Further overriding of the threshold, that is growing of the initial population inversion, results in the ratio of the trailing edge duration to the leading edge duration becomes greater. At large times the incoherent radiation dominates. 

 Because very small value of the initial population inversion can provide generation of pulses of stimulated radiation, it is of interest to determine the shape of these pulses for different values of the initial population inversion levels and when the field energy absorption should be taken into account.  These pulses can be easily detected in experiments. In addition, after experimental validation of this model, it will be possible to use these approaches for analysis of the cosmic radiation that might help explain such abundance of coherent radiation sources in space. 

 In this paper, we study the characteristics of the pulses of stimulated (coherent) radiation as a function of the initial inversion and absorption level in the system. The dynamics of the emission process in the simplified model is compared with the dynamics of change in the number of quanta in the traditional model, where the separation into spontaneous and stimulated components is not carried out.

 If there is exist a recovery mechanism for the population inversion, the pulsating mode of stimulated emission generation becomes possible. The integral radiation intensity at this may be increased several times. This approach can be used for analysis of the cosmic radiation that might help explain a great variety of pulsating radiation sources in space. In present work, we investigate the characteristics of the periodic pulse generation depending on the initial inversion, the pumping level and the absorption rate.

\section{Traditional description of two-level system}

Following to A. Einstein \cite{einstein1916quantentheorie} let describe a two-level system with transition frequency $\varepsilon _{2} -\varepsilon _{1} =\hbar \omega _{12} $ by the following set of equations: 
\begin{align}
	&\partial n_{2} /\partial t=+w_{12} N_{k} n_{1}-(u_{21} +w_{21} N_{k} ) n_{2}, \nonumber \\
	&\partial n_{1} /\partial t=-w_{12}  N_{k}  n_{1} +(u_{21} +w_{21}  N_{k} ) n_{2},
	\label{eq1} 
\end{align}
where the sum of level populations $n_{1} +n_{2} =N$ remains constant, $u_{21} n_{2} $ is the rate of change in the number density of atoms due to spontaneous emission. The rates of change in level population due to stimulated emission and absorption are $w_{21} N_{k} n_{2} $ and $w_{12} N_{k} n_{1} $ correspondingly. The number of quanta $N_{k} $ on the transition frequency $\omega _{k} $ is governed by the equation 
\begin{equation}
	\frac{\partial N_{k} }{\partial t} =(u_{21} +w_{21} N_{k} ) n_{2} -(w_{12} N_{k} ) n_{1}.  
	\label{eq2}
\end{equation}

Note that in 3D case, the relation between coefficients $u_{21}$ and $w_{21}$ is following
\begin{equation}
\frac{u_{21}} {w_{21}} = \alpha = \frac{1}{\hbar\omega}\frac{A_{21}}{B_{21}} = \frac{2\omega^2}{\pi c^3}, \nonumber
\end{equation}
where $A_{21}$ and $B_{21}$ are corresponding Einstein's coefficients. The yellow spectrum line gives $\alpha \approx 0.25$ and the violet edge of the visible spectrum gives $\alpha \approx 0.6$. 

The losses of energy in active media are caused mainly by radiation outcome from a resonator. These radiative losses can be calculated by imposing the correct boundary conditions on the field. Thus, they can be estimated in rather common form with the following parameter:
\begin{multline}
	\delta =\oiint_{S}\frac{\partial \omega }{\partial \vec{k}}  \frac{1}{4\pi } \vec{E}\times \vec{H}ds \left/{\oiiint _{V}\frac{\partial (\omega \varepsilon (\omega ,\vec{k}))}{\partial \omega } }\right. \times \\ 
	\times \frac{1}{8\pi } (|\vec{E}|^{2} +|\vec{H}|^{2} )dv,
	\label{eq3}
\end{multline}
i.e. as the ratio of the energy flow passing through the resonator mirrors to the total field energy within resonator. It is important, that the characteristic size of the resonator $L$ should be much less than the characteristic time of field variation $\tau \sim |\vec{E}|^{2} (\partial |\vec{E}|^{2} /\partial t)^{-1} $ multiplied by the group velocity of oscillations $|\partial \omega /\partial \vec{k}|$.  In this case the radiative losses through the mirrors can be replaces by distributed losses within the resonator volume. The threshold of instability leading to exponential growth of stimulated emission in this case is defined by condition $\mu _{0} >\mu _{TH1} $ (see, for example [6], where                    
\begin{equation}  
	\mu _{TH1} =\delta /w_{21}.
\label{eq4}                                                  
\end{equation} 
Equations \eqref{eq1}-\eqref{eq2} can be rewritten in the form
\begin{align} 
	&\partial n_{2} /\partial \tau =-\alpha n_{2} -\mu  N_{k},
	\label{eq5}\\
	&\partial \mu /\partial \tau =-2\alpha n_{2} -2\mu N_{k},
	\label{eq6}\\
	&\partial N_{k} /\partial \tau =\alpha n_{2} +\mu  N_{k},
\label{eq7}
\end{align}
where $\tau =w_{21} t$, $u_{21} =w_{21} =w_{12} $. Since the purpose of this work is to find the threshold of the initial population inversion, which starts the exponential growth of the number of emitted quanta, we will restrict our consideration by the case $\mu =n_{2} -n_{1} \ll n_{1} ,n_{2} $. It follows from Eqs. \eqref{eq5}-\eqref{eq7} that $N_{k} =N_{k0} +(\mu _{0} -\mu )/2\approx (\mu _{0} -\mu )/2$, and at large times $n_{2st} \approx N/2=-\mu _{st} (\mu _{0} -\mu _{st} )/2$, where $\mu _{0} =\mu (\tau =0)$, $N_{k0} =N_{k} (\tau =0)$. Hence, we find the stationary value of the inversion 
\begin{equation}
	\mu _{st} =(\mu _{0} /2)-\sqrt{(\mu _{0} /2)^{2} +N}.
\label{eq8} 
\end{equation}

Two cases are of interest. When the initial population inversion is sufficiently large $(\mu _{0} /2)^{2} \gg N$, it rapidly decreases to its steady-state value $\mu \to \mu _{st1} =-(N/\mu _{0} )$ with $|\mu _{st1} |\ll\mu _{0} $. The number of quanta at this growths exponentially and asymptotically tends to a stationary level $N_{k} \to N_{kst1} =\mu _{0} /2$. It is obviously that in this case the stimulated emission dominates (the second terms in r.h.s. of Eqs. \eqref{eq5}-\eqref{eq7}.

The second case of interest corresponds to relatively small initial inversion $(\mu _{0} /2)^{2} \ll N$. Here, $\mu $ tends to its stationary value $\mu \to \mu _{st} =-(N)^{1/2} $, where $|\mu _{st} |>\mu _{0} $, and the number of quanta reaches the limit $N_{k} \to N_{kst2} =N^{1/2} $.  

If the spontaneous emission only dominated (the first terms on the r.h.s. of Eqs. \eqref{eq5}-\eqref{eq7}, the characteristic time to reach the steady-state number of photons will be of the order of $\Delta \tau \sim \tau _{m} =\mu _{0} /N>\mu _{0} ^{-1} $ in the first case and $\Delta \tau \sim 1/\sqrt{N} <\mu _{0} ^{-1} $ in the second case, where $\mu _{0} ^{-1} $ is the characteristic time of exponential growth of the number of photons in the first case. This means that the exponential growth of the number of photons in the second case is suppressed and the role of the second term in r.h.s. of Eqs. \eqref{eq5}-\eqref{eq6} comes to stabilize the number of particles and the inversion level due to the absorption process. 

Thus, it is clear that the scenario of the process changes, if the initial value of the inversion $\mu _{0} $ is more or less than a threshold value [5]: 
\begin{equation}
	\mu _{TH2} =2\sqrt{N}.
\label{eq9} 
\end{equation}

The suppression of the exponential growth of the number of photons when $\mu _{0} <\mu _{TH2} =2\sqrt{N}$  demonstrates not only the changes in scenario of the process, but it suggests that the stimulated emission is suppressed by preferential growth of spontaneous emission. Indeed, the first term in  r.h.s. of Eqs. \eqref{eq5}-\eqref{eq7}, which is responsible for the spontaneous emission, reduces inversion to zero in a very short time $\tau <1/\mu _{TH2} $, thus excluding the possibility of exponential growth of the number of photons, which is characteristic for the induced processes.

It is useful, at least qualitatively, to examine the nature of changes in emission characteristics of an inverted system near the threshold $\mu _{TH2} $. It should be expected also other specific features in the radiation nature, including the formation of a short pulse of coherent radiation against the background of incoherent field \cite{kuklin2012realization}. 


\section{Qualitative model of two-level system}

First of all, in order to understand the further, it should be remembered that the oscillator emits under the action of an external coherent field with the same frequency and phase as the stimulating field, that is, the external radiation and radiation of the oscillator stimulated by it occur to be coherent \cite{birnbaum1965optical}, \cite{blotmbergen1965nonlinear}. Moreover, the greater intensity of the coherent component of the external field, the more energy the oscillator loses per unit time by radiation. On the other side, the spontaneous emission is the process independent of the external radiation field and incoherent, at least for a uniform distribution of emitters. 

Neglecting the stage when the number of photons is saturated, we can at least qualitatively assume that the terms in r.h.s. of Eqs. \eqref{eq1}-\eqref{eq2} proportional to $N_{k}$ correspond to the stimulated processes, as well as the photons which number $N_{k} $ is incorporated in these terms will be assumed coherent. With these general principles in mind, we expand the total number of photons into two components $N_{k} =N_{k} ^{(incoh)} +N_{k} ^{(coh)}$ where $N_{k} ^{(incoh)}$ and $N_{k} ^{(coh)}$ are the numbers of quanta of spontaneous and stimulated radiation correspondingly. Then Eqs. \eqref{eq2}-\eqref{eq3} take the form 
\begin{align} 
	&{\partial n_{2}}/{\partial t} = w_{12} N_{k} ^{(coh)} n_{1} - \left[u_{21} +w_{21} N_{k} ^{(coh)} \right] n_{2},
	\label{eq10}\\
	&{\partial n_{1}}/{\partial t} = -w_{12} N_{k} ^{(coh)} n_{1} +\left[u_{21} +w_{21} N_{k} ^{(coh)}\right]n_{2},
	\label{eq11}\\
	&{\partial N_{k} ^{(incoh)}}/{\partial t} = u_{21} n_{2},
	\label{eq12}\\
	&{\partial N_{k} ^{(coh)}}/{\partial t} = w_{21} N_{k} n_{2} - w_{12} N_{k} n_{1}.
	\label{eq13}
\end{align}
Assuming $u_{21} =w_{21} =w_{12}$ and $n_{2} =(N+\mu )/2$, we obtain                                      
\begin{align} 
	&\partial n_{2} /\partial \tau =-n_{2} -\mu N_{k} ^{(coh)},
	\label{eq14}\\
	&\partial \mu /\partial \tau =-2n_{2} -2\mu  N_{k} ^{(coh)},
	\label{eq15}\\
	&\partial N_{k} ^{(incoh)} /\partial \tau =n_{2},
	\label{eq16}\\
	&\partial N_{k} ^{(coh)} /\partial \tau =\mu  N_{k} ^{(coh)}.
\label{eq17}
\end{align}
where $N=n_{1} +n_{2} $ is a total number of emitters.

Let compare the dynamics of the processes described by Eqs. \eqref{eq14}-\eqref{eq17} and by Eqs. \eqref{eq5}-\eqref{eq7}. In order to do this, we represent them as follows:
\vspace{0.2cm}
\paragraph{The modeling set of equations with separation of photons into coherent and incoherent sorts}
\begin{align} 
	&\partial {\rm M} /\partial T=-N_{0} -2{\rm M} \cdot {\rm N} _{c},
	\label{eq18}\\
	&\partial {\rm N} _{inc} /\partial T=N_{0} -\theta \cdot {\rm N} _{inc},
	\label{eq19}\\
	&\partial {\rm N} _{c} /\partial T={\rm M} \cdot{\rm N} _{c} -\theta \cdot {\rm N} _{c},
	\label{eq20}
\end{align}

\paragraph{Traditional set of equations}
\begin{align} 
	&\partial {\rm M} _{1} /\partial T=-N_{0} -2{\rm M} _{1} \cdot {\rm N} _{1},
	\label{eq21}\\
	&\partial {\rm N} _{1} /\partial T={N_{0}}/{2} +{\rm M} _{1} \cdot {\rm N} _{1} -\theta \cdot {\rm N} _{1},
	\label{eq22}
\end{align}
where ${\rm N} _{inc} =N_{k} ^{(incoh)} /\mu _{0} $, ${\rm N} _{c} =N_{k} ^{(coh)} /\mu _{0} $, ${\rm M} =\mu /\mu _{0} $, ${\rm M} ={\rm M} _{1} =\mu /\mu _{0} $ $T=w_{21} \cdot \mu _{0} \cdot t=\mu _{0} \cdot \tau $       ${\rm N} _{1} =N_{k} /\mu _{0} $. The only free parameter convenient for the analysis is $N_{0} =N/\mu _{0}^{2} $. For correct comparison, we assume that the total number of real states is $N=n_{1} +n_{2} =10^{12} $, and the threshold inversion is $\mu _{0th} =\sqrt{N} =10^{6} $. Transition to a unified time scale will be carried out as follows $T=\tau \cdot \mu _{0} $, where $T$is time for each case. 

Let choose the following initial values:
\begin{align*}
	&{\rm M} (T=0)={\rm M} _{1} (T=0)=1, \\ 
	&{\rm N} _{inc} (T=0)={\rm N} _{inc} /\mu _{0} =3\cdot 10^{4} /\mu _{0},\\
	&{\rm N} _{c} (T=0)={\rm N} _{c} /\mu _{0} =3\cdot 10^{4} /\mu _{0},\\
	&{\rm N} _{1} (T=0)={\rm N} _{k} /\mu _{0} =3\cdot 10^{4} /\mu _{0}.
\end{align*}
The radiation losses are taking into account by the term $\theta =\delta /\mu _{0} $, where $\delta $ is defined in Eq.~\eqref{eq3}.

\begin{figure}[t] \centering
	\includegraphics[width=0.47\textwidth]{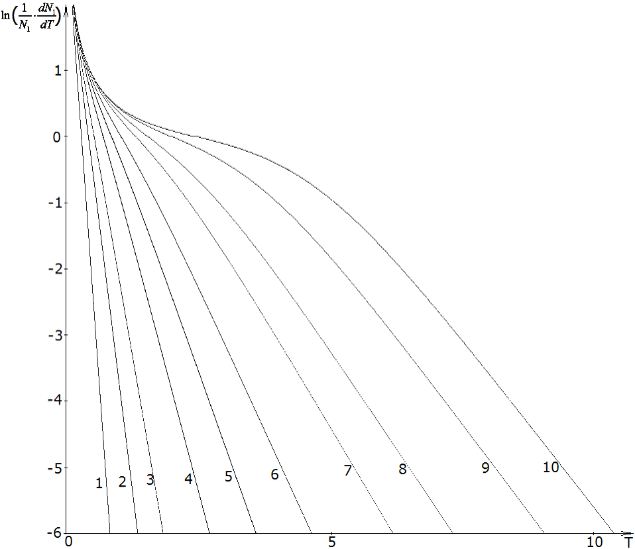}
	\caption{Evolution of the value $\ln ({N_{1}^{-1}dN_{1} / {dT}})$ for different values of the parameter $N_{0} =(n_{1} +n_{2} )/(n_{2} -n_{1} )^{2} $:   1) $N_{0} =30$; 2)~$N_{0} =10$; 3)~$N_{0} =5$; 4)~$N_{0} =2$; 5)~$N_{0} =1$; 6)~$N_{0} =0.5$; 7)~$N_{0} =0.2$; 8)~$N_{0} =0.1$; 9)~$N_{0} =0.03$.}
	\label{fig1}
\end{figure}

\begin{figure*}[t] \centering
	\begin{subfigure}[b]{0.47\textwidth}
		\includegraphics[width=\textwidth]{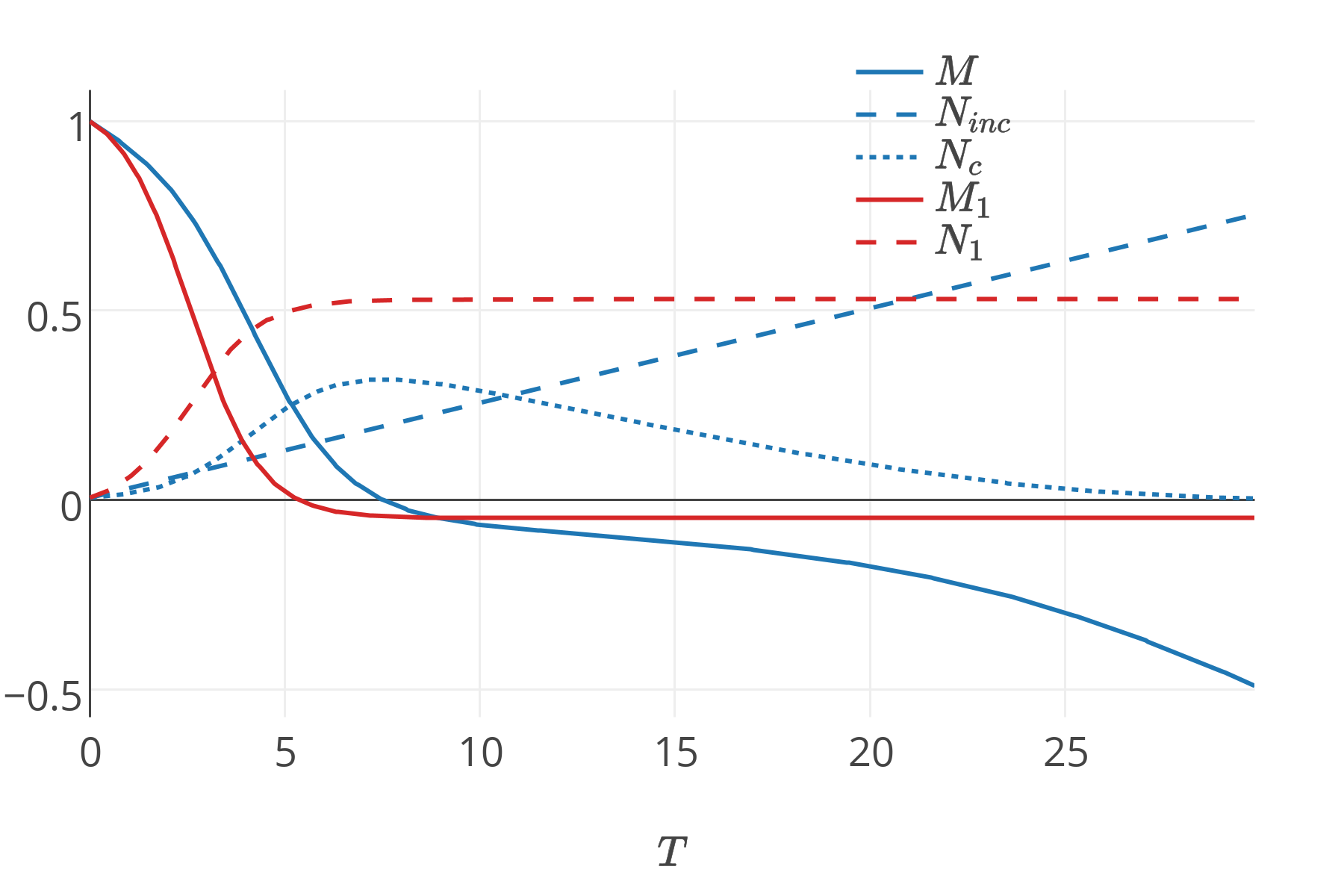}
		\caption{$\theta =0$, $N_{0} =N/\mu _{0}^{2} =0.05$}
	\end{subfigure}
	\begin{subfigure}[b]{0.47\textwidth}
		\includegraphics[width=\textwidth]{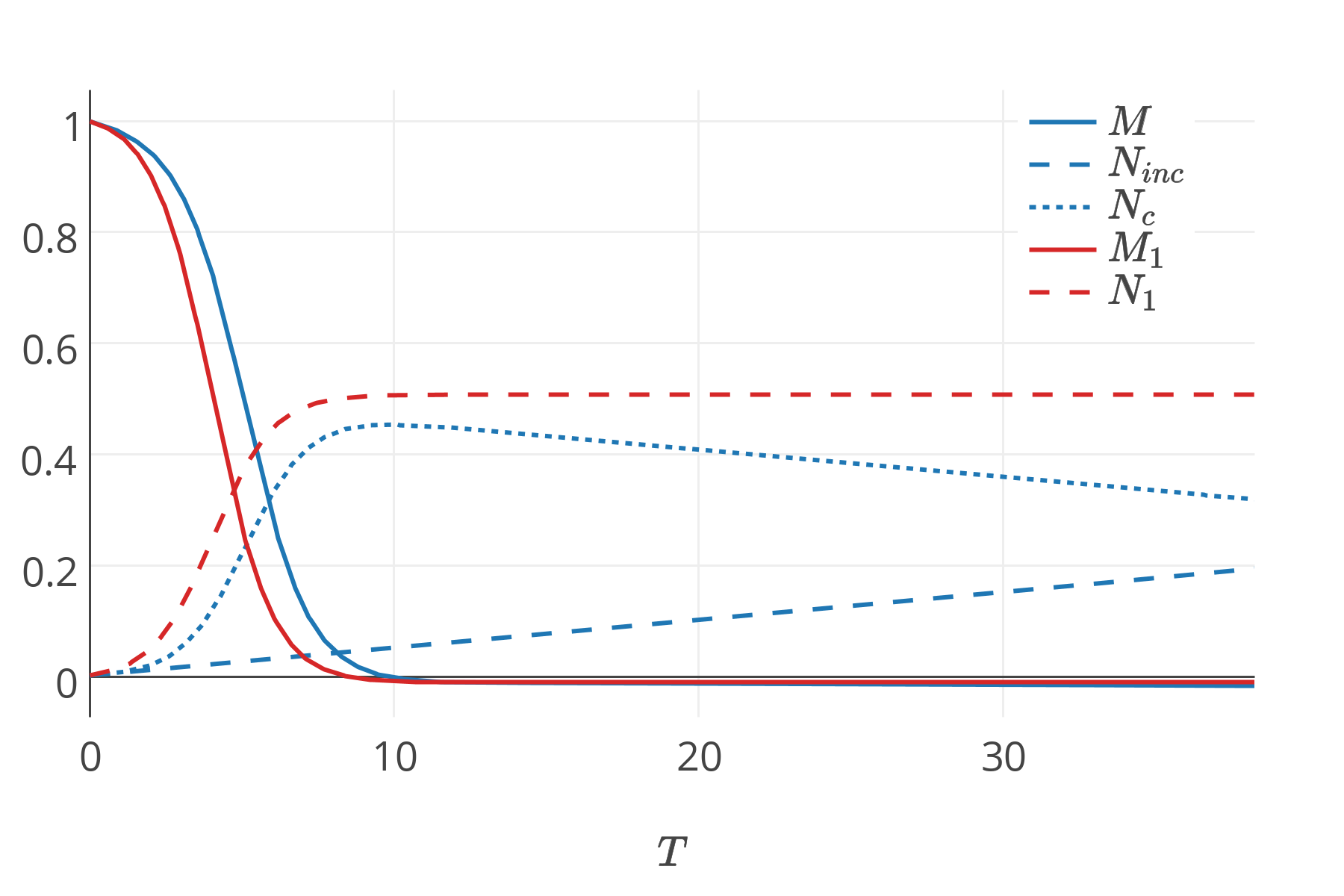}
		\caption{$\theta =0$, $N_{0} =N/\mu _{0}^{2} =0.01$}
	\end{subfigure}
	
	\caption{Evolution of the inversion ${\rm M} _{1} $ and the number of photons ${\rm N} _{1}$ (red lines), calculated from the traditional model, Eqs.~\eqref{eq21}-\eqref{eq22} and the inversion ${\rm M} $, the number of stimulated (coherent) photons ${\rm N} _{c} $ and also the number of spontaneous (incoherent) photons ${\rm N} _{inc}$ (blue lines), calculated from the model with "separated" photons, Eqs.~\eqref{eq18}-\eqref{eq20} in the lossless case $\theta=0$.}
	\label{fig2_3}
\end{figure*}

Figure 1 demonstrates a change in dynamics of the process with increase in the starting population inversion \eqref{eq9} simulated by Eqs. \eqref{eq21} - \eqref{eq22}, where $N_{0} \subset (30\div 0.01)$.

Let discuss the reasons why it makes sense to use a qualitative system of equations \eqref{eq18} - \eqref{eq20} near the threshold \eqref{eq9}.

Within the framework of the classical description, the total intensity of the spontaneous emission of an ensemble of particles-oscillators, whose phases are distributed randomly and uniformly, can be found as a sum of individual intensities produced by each particle-oscillator being in an excited state. As for the stimulated emission, the radiation field strength is so great that synchronizes the phase both of the emitting and absorbing oscillators. Thus, the sign of the population inversion $\mu =n_{2} -n_{1} $ determines is the stimulated field will increase or decrease. Note, that the characteristic time of this process is inversely proportional to $\mu $. However, if the coherent field is absent, the oscillators in the excited state will emit only spontaneously, because their phases are not synchronized. The absorption of the spontaneous field by the unexcited particles-oscillators can be ignored since they are placed in a random rapidly alternating field, which averaged effect is negligible.

In the quantum case, the traditional model \eqref{eq14} - \eqref{eq17}  includes the term $\mu N_{k} $, which is responsible for the stimulated processes of excitation and absorption. But it has no physical meaning below the threshold \eqref{eq9}, since in this case there is no an intense stimulated field, which is able to synchronize the emission of many emitters.

The attention should be given to a change in the rate of emitted quanta with crossing of the threshold \eqref{eq9}. For greater values of the initial inversion, the stimulated emission begins to prevail and the regime of exponential growth in the number of quanta becomes more pronounced. 

In the absence of radiative losses, the simulation of Eqs. \eqref{eq18} - \eqref{eq20} shows that after the coherent pulse drops, the spontaneous emission continues to increase. Within framework of the traditional model \eqref{eq21} - \eqref{eq22}, absorption restricts the growth of the number of quanta and radiation intensity tends to a stationary level [9].

However, comparing the dynamics of the processes it can be understood that after the amplitude of the stimulated coherent pulse decreases, the spontaneous emission becames dominant. That is, at times exceeding the duration of the coherent pulse the spontaneous incoherent radiation prevails.

The absorption of photons suppresses the generation, so we choose relatively low level of energy loss, that is $\delta =2\cdot 10^{5} $ and $\delta =4\cdot 10^{5} $. The generation process in this case keeps the same features, but the absorption limits the lifetime of the generation and the differences between two models are less pronounced.

\begin{figure*}[t] \centering
	\begin{subfigure}[b]{0.47\textwidth}
		\includegraphics[width=\textwidth]{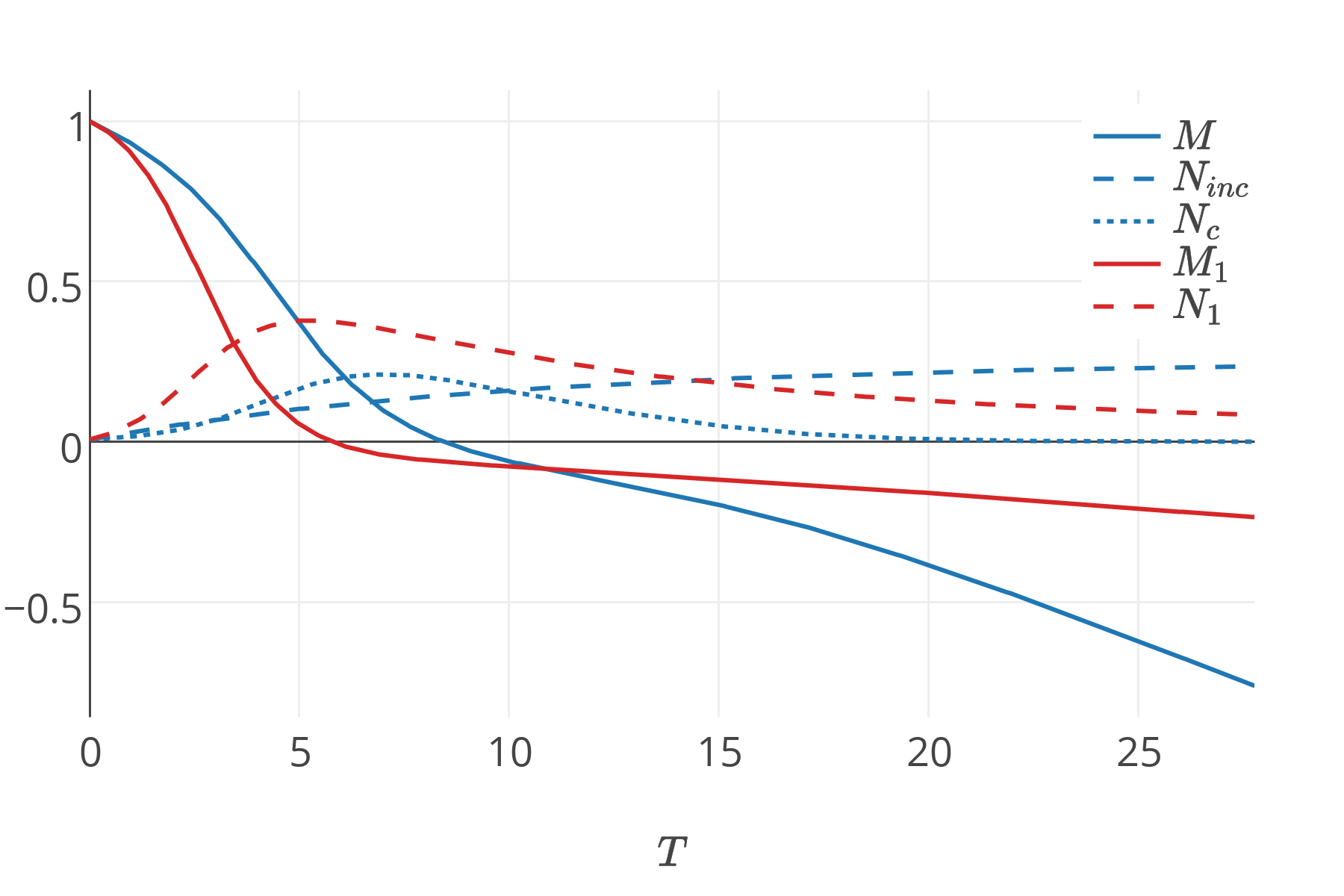}
		\caption{$\delta =2\cdot 10^{5} $, $\theta =\delta /\mu _{0} $=0.045 and $N_{0} =N/\mu _{0}^{2} =0.05$}
	\end{subfigure}
	\begin{subfigure}[b]{0.47\textwidth}
		\includegraphics[width=\textwidth]{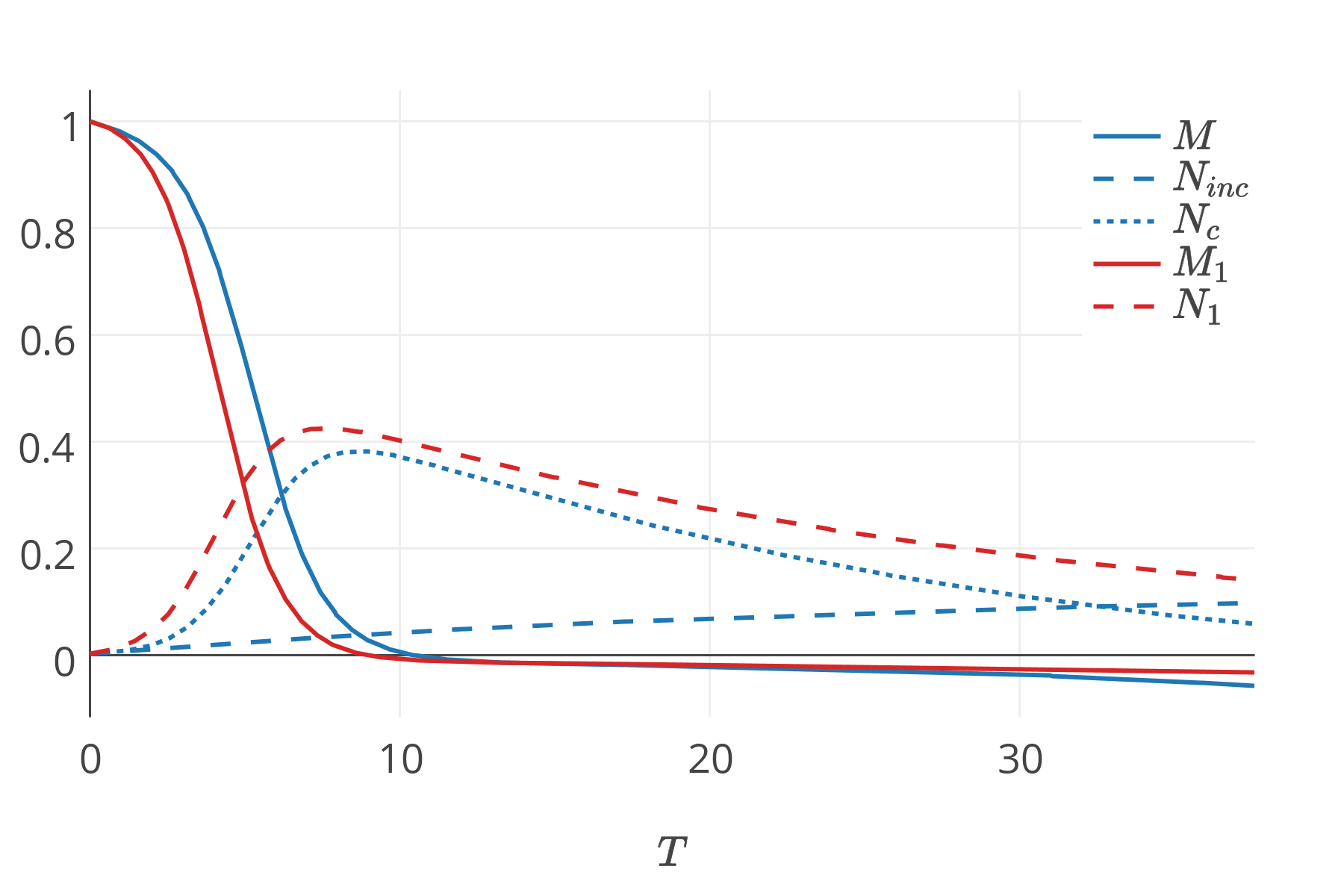}
		\caption{$\delta =4\cdot 10^{5} $, $\theta =\delta /\mu _{0} $=0.04 and $N_{0} =N/\mu _{0}^{2} =0.01$}
	\end{subfigure}
	
	\caption{Evolution of the inversion ${\rm M} _{1} $ and the number of photons ${\rm N} _{1}$ (red lines), calculated from the traditional model, Eqs.~\eqref{eq21}-\eqref{eq22} (a) and the inversion ${\rm M} $, the number of stimulated (coherent) photons ${\rm N} _{c} $ and also the number of spontaneous (incoherent) photons ${\rm N} _{inc}$ (blue lines), calculated from the model with "separated" photons, Eqs.~\eqref{eq18}-\eqref{eq20} in the dissipation case $\theta\ne 0$.}
	\label{fig4_5}
\end{figure*}

Now, let discuss the quantitative characteristics of the coherent pulse. Figures 6 and 7 demonstrate the shape of the coherent pulse in lossless case and in presence of absorption for different initial value of the population inversion.
\begin{figure*}[t] \centering
	\begin{subfigure}[b]{0.47\textwidth}
		\includegraphics[width=\textwidth]{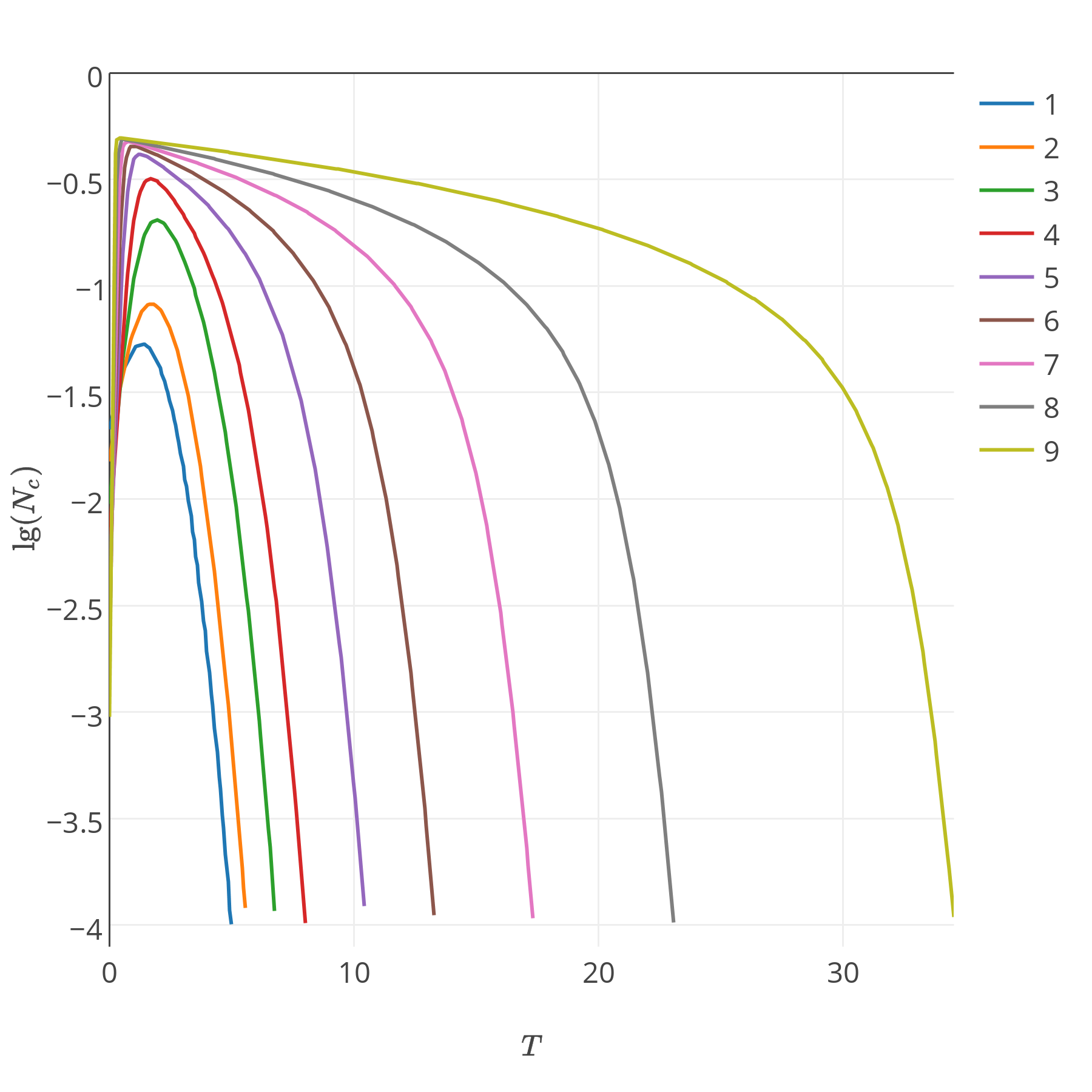}
		\caption{the lossless case ($\theta =0$)}
	\end{subfigure}
	\begin{subfigure}[b]{0.47\textwidth}
		\includegraphics[width=\textwidth]{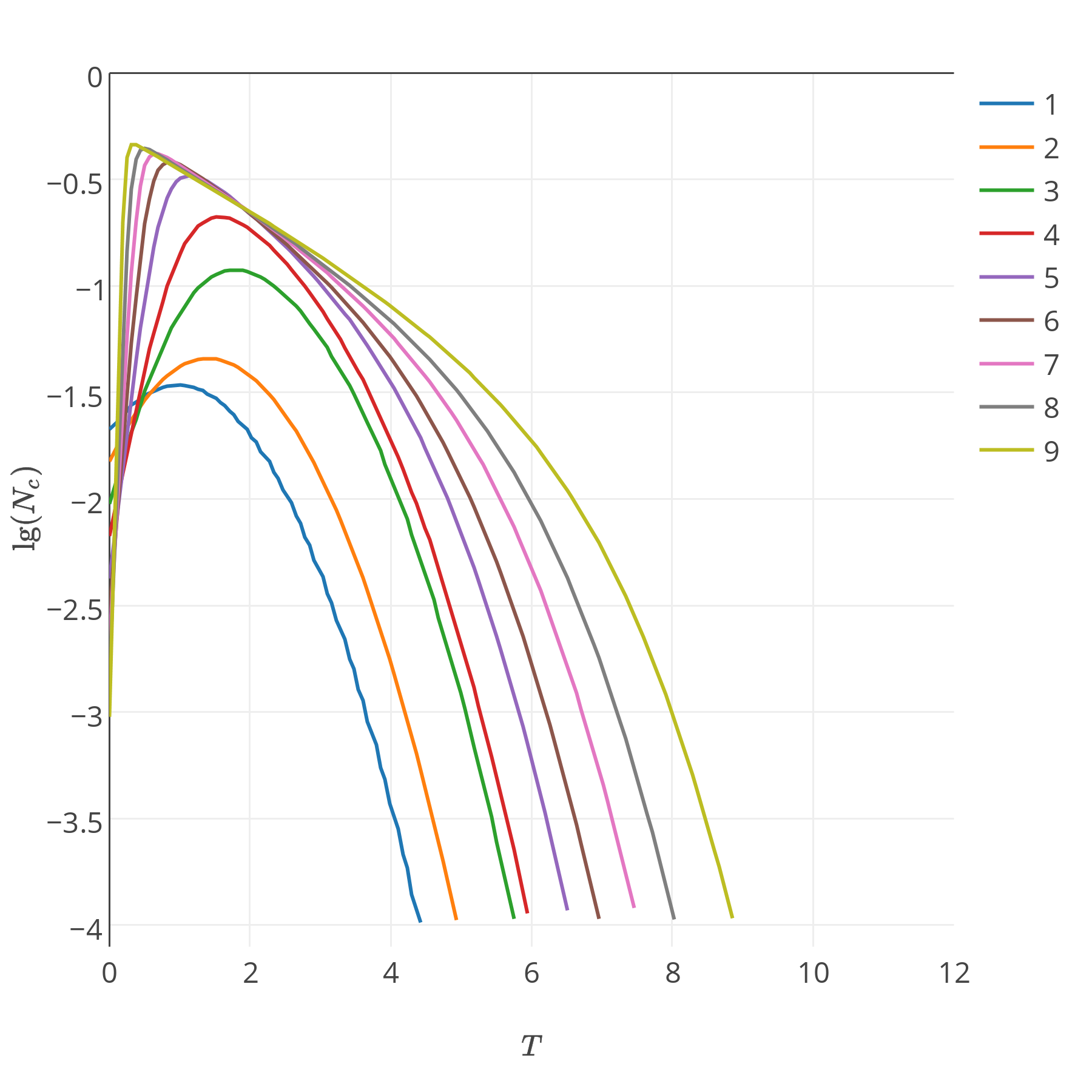}
		\caption{the dissipation case ($\delta =4\cdot 10^{5} $)}
	\end{subfigure}
	
	\caption{Evolution of coherent pulse shape for different values of initial population inversion:\\ 1)$\mu _{0} =\sqrt{2} \cdot 10^{6} $; 2)~$\mu _{0} =2\cdot 10^{6} $; 3)$\mu _{0} =\sqrt{10} \cdot 10^{6} $; 4) $\mu _{0} =\sqrt{20} \cdot 10^{6} $; 5) $\mu _{0} =\sqrt{50} \cdot 10^{6} $; 6) $\mu _{0} =10^{7} $; 7) $\mu _{0} =\sqrt{2} \cdot 10^{7} $; 8)~$\mu _{0} =2\cdot 10^{7} $; 9)$\mu _{0} =\sqrt{10} \cdot 10^{7} $}
	\label{fig6_7}
\end{figure*}

\section{On the nature of pulsating radiation}
The equations \eqref{eq1} with consideration of the collisional mechanism of inversion generation can be rewritten as follows: 
\begin{align} 
	&\partial n_{2} /\partial \tau =-n_{2} -\mu N_{k} +\frac{\nu }{u_{21} } n_{1},  \label{eq23} \\
	&\partial n_{1} /\partial \tau =+n_{2} +\mu N_{k} -\frac{\nu }{u_{21} } n_{1}, \label{eq24}
\end{align}
where $\nu $ is the effective frequency of collisions, which kick a particle from lower energy level to upper one. Under equilibrium condition, it is obvious that $\nu \sim u_{21} $ and Eq. \eqref{eq18} takes the form 
\begin{equation}  
\partial \mu /\partial \tau =(\nu -u_{21} )n_{1} -2\mu -2\mu \cdot N_{k},
\label{eq25} 
\end{equation} 
where $(\nu -u_{21} )n_{1} =\mu _{0}^{2} I_{0} $.


\begin{figure}[b] \centering
	\includegraphics[width=0.47\textwidth]{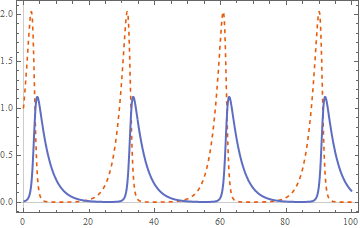}
	\caption{The repetition pulse train, resulting as a solution of Eqs. \eqref{eq27}-\eqref{eq29} for $\tilde{{\rm \Gamma }}=0.1$ and $\theta=0.4$.}
	\label{fig8}
\end{figure}


Near the threshold \eqref{eq9} when $n_2 \approx n_1$, it is reasonable to use namely a qualitative system of equations \eqref{eq18}-\eqref{eq20}. Then the steady state of the spontaneous emission is determined by the value $\delta N_{k} ^{(incoh)} \simeq N/2$, but the energy flow of the stimulated emission is equal to $\delta N_{k} ^{(coh)} $. In order to describe the behavior of a two-level system in presence of the radiation losses and taking into account collisions, Eqs. \eqref{eq18} - \eqref{eq20} can be rewritten in a convenient form
\begin{align}  
	&\partial {\rm M} /\partial T=I_{0} -2{\rm M} -2{\rm M} \cdot{\rm N} _{c} ,
	\label{eq26}\\ 
	&\partial {\rm N} _{inc} /\partial T=N_{0} /2-\theta \cdot{\rm N}_{inc},
	\label{eq27}\\                                      
	&\partial {\rm N} _{c} /\partial T={\rm M} \cdot{\rm N} _{c} -\theta \cdot{\rm N} _{c}.  \label{eq28}                                          
\end{align}

It is reasonable to consider the case of $\mu_0=\delta/w_{21}$. Then both the threshold (\ref{eq4})  and (\ref{eq9}) are nearly equal
\begin{equation*}
\mu_{TH2}\approx\mu_{TH1}=\delta/w_{21}.
\end{equation*} 
 
In the case, when $\mu >\mu _{TH2} $ and $M(0)>\theta $, the relaxation oscillations appear in the system resulting in a stationary state ${\rm N} _{cst} \approx 2+I_{0} /2\theta $, $M_{st} =\theta $. The total radiation flow outside the system in assumed terms is equal to
\begin{equation*}
\theta \cdot {\rm N} _{cst} +\theta \cdot {\rm N} _{incst} =(I_{0} +2\theta )/2+N_{0} /2\ge N_{0} /2. 
\end{equation*} 

Note, that in presence of an external mechanism, which provides an exceeding of the inversion over its stationary value $M_{st} =\theta $, the Eq.~\eqref{eq14} can be supplemented by the driving term
\begin{equation}
\partial {\rm M} /\partial T=\Gamma {\rm M} -2{\rm M} -2{\rm M} \cdot {\rm N} _{c} +2I_{0}.
\label{eq29} 
\end{equation}

As will be shown bellow, the continuous oscillations can arise only when $\Gamma>2$. One of the possible physical mechanisms to provide the sufficient pumping can be a convection
\begin{equation}
-v\frac{\partial M}{\partial X} \approx v\frac{M}{L}=\Gamma M >0, 
\label{eq29a} 
\end{equation}
where $v\nabla M \approx ML^{-1}$ is the convective transport of the population inversion.

The equations \eqref{eq26}-\eqref{eq28} are similar to so-called Statz-DeMars equations \cite{statz1960transients}, which  describe the relaxation oscillations in a two-level media in the presence of the pump and energy losses. The only difference in equation \eqref{eq29} is the first term in r.h.s. that provides the generation of the population inversion. Namely this term changes the characteristics of pulse generation from relaxation to periodic. 

The equations \eqref{eq27}-\eqref{eq29} have a solution in a form of periodical sequence of coherent pulses (see Fig.1) against a background of the mean radiation flow, if $\nu =u_{21} $:
\begin{equation} 
\theta \cdot {\rm N} _{cst} +\theta \cdot {\rm N} _{incst} =(\tilde{{\rm \Gamma }}\theta +N_{0} )/2,
\label{eq30}               
\end{equation} 
where $\tilde{\Gamma }=\Gamma -2>0$, $\rm N_{cst}\approx(I_0+\tilde{\Gamma }\theta)/2\theta$, $\tilde{\Gamma }>I_0/\theta$. The pulse repetition rate is $\sqrt{\theta \cdot \tilde{\Gamma }} $. The integral radiation intensity on the pulse peak can exceed the background value in several times.

It should be noted that the radiation losses of the field energy $\theta $ in open systems is defined as the ratio of the energy flux from the object to the energy in its volume, and therefore this parameter decreases with increase of the radius of the system $R$ as $c/R$, where $c$ is the speed of light. This means that an increase in size $R$ reduces the losses $\theta $, which in turn, provides a higher intensity of the stimulated emission. That is, at the same parameters of the system, the larger objects should generate more intense pulses but with less repetition rate.

\section{Conclusion}

\noindent The threshold of coherent emission generation, discussed in this paper, corresponds to the case when the intensity of spontaneous and stimulated coherent radiation become equal. The stimulated emission in this case can be considered as completely coherent or as a set of narrow wave packets of coherent radiation. When the initial population inversion crosses the threshold \eqref{eq9}, the process of generation undergoes qualitative changes. The excess of the threshold \eqref{eq9} leads to an exponential growth in the number of quanta. If we make the assumption that the stimulated emission is mainly coherent, the nature of this threshold can be explained as follows: generation of coherent radiation begins only after crossing of this threshold. In this work, we have tried to develop a qualitative model of this process. 

It follows from results of numerical simulation that the number of coherent quanta tends to $\mu _{0} /2$ with increase of the initial population inversion in agree with the theory of superradiance. If we fix all parameters except the inversion, the duration of the coherent pulse estimated by its half-width significantly increases with increasing initial inversion in the absence of absorption. The foregoing estimates of the characteristic times of the process are confirmed by numerical calculations. For relatively small inversion levels $\sqrt{N} \ll \mu _{0} \ll N$ the coherent emission is always presents as a rather short pulse with duration of $\tau \sim (\mu _{0} /N)$. At large times $\tau >(\mu _{0} /N)$ the incoherent radiation dominates.  Since the model \eqref{eq18}-\eqref{eq20} doesn't take into account absorption of the incoherent radiation, it becomes inapplicable after this time. It is important to note that the time when the total number of photons reaches the steady state in the model \eqref{eq21}-\eqref{eq22} after exceeding the threshold \eqref{eq9} is comparable with the time when the number of spontaneous photons achieves the same values $\Delta \tau \sim \tau _{m} =\mu _{0} /N$ in the model \eqref{eq18}-\eqref{eq20}. 

If the absorption is taken into account, even a small, the coherent pulse duration remains almost unchanged with an increase in the population inversion, at least far enough above the threshold. The ratio of the pulse-trailing edge duration to the pulse-leading edge duration (the latter, by the way, is inversely proportional to the initial inversion) is growing with increasing of the initial inversion.  The duplication of absorption reduces the pulse duration by half. In an absorbing medium and when with significant overriding of the threshold \eqref{eq9}, the difference between the traditional model and our qualitative description become insignificant.

In presence of a mechanism for supporting the population inversion, such as the convection from optically dense layers into less dense one, transparent for radiation, the first term in r.h.s. of Eq. \eqref{eq29} becomes positively defined. Then one can obtain the periodic train of stimulated radiation pulses as a solution of Eqs. \eqref{eq27}-\eqref{eq29}. In the case of convection, $\Gamma \approx v/L$ in Eq. \eqref{eq29}, where $v$ is a characteristic speed of the convection and $L$ is a characteristic scale of medium density gradient.

The field intensity of considered pulses occurs of the same order or more than the intensity of the spontaneous emission, because of even for small population inversion $\mu =(n_{2} -n_{1} )\ll N=(n_{1} +n_{2} )<\mu ^{2} =(n_{2} -n_{1} )^{2} $ the intensity of the spontaneous emission is proportional to $N/2$ near the threshold \eqref{eq2}, but the intensity of the stimulated emission is proportional to $\mu ^{2} =(n_{2} -n_{1} )^{2} \ge N/2$. 

Thus, this mechanism may be responsible for periodic changes in luminosity of physical systems with low population inversion of higher energy levels. In particular, such a mechanism could explain the periodic variation of the luminosity of stars with intensive convective flows in their atmosphere.


\bibliography{maser} 
\end{document}